Corrigendum to "Is the expertise of evaluation panels congruent with the research interests of the research groups: a quantitative approach based on barycenters" [Journal of Informetrics 9(4) (2015) 704-721]

A. I. M. Jakaria Rahman [a], Raf Guns[a], Ronald Rousseau[b,c], Tim C. E. Engels[a,d]

[a] Centre for R&D Monitoring (ECOOM), Faculty of Social Sciences, University of Antwerp, Middelheimlaan 1, B-2020 Antwerp, Belgium

[b] KU Leuven, Dept. of Mathematics, Celestijnenlaan 200B, B-3001 Leuven, Belgium

[c] University of Antwerp, Faculty of Social Sciences, B-2000 Antwerp, Belgium

[d] Antwerp Maritime Academy, Noordkasteel Oost 6, B-2030 Antwerp, Belgium

In Rahman, Guns, Rousseau, and Engels (2015) we investigated, among other things, the following research question: 'How can we quantify the overlap of expertise between two entities (e.g., a research group and a panel) using publication data?'. In order to answer this research question we considered two approaches: one based on barycenters in two dimensions, and another one based on so-called "barycenters" in N dimensions, where N denotes the total number of Web of Science Subject Categories (WoS SCs). The first of these approaches uses overlay maps (Rafols, Porter, & Leydesdorff, 2010). Each SC has a place on this map, characterized by the corresponding coordinates, denoted as $(L_{j,1}, L_{j,2})$, $j = 1, …, N$. Now for each panel member and for each research group a barycenter is calculated and Euclidean distances between barycenters can be determined. Coordinates of these barycenters (in 2 dimensions) are given as

$$C_1 = \frac{\sum_{j=1}^{N} m_j L_{j,1}}{T} \; ; \; C_2 = \frac{\sum_{j=1}^{N} m_j L_{j,2}}{T} \qquad (1)$$

where $m_j$ is the number of publications of the unit under investigation (panel member, research group) belonging to category $j$; this category $j$ has coordinates $(L_{j,1}, L_{j,2})$ in the base map. The total number of publications of the unit under investigation is denoted as $T = \sum_{j=1}^{N} m_j$. This approach is the barycenter approach as announced in the title of Rahman et al. (2015). In this way distances between entities, as represented by their barycenters, can be calculated leading to quantitative results answering the aforementioned research question.

Urged by a colleague, we point out that the term 'barycenter' taken on its own, has no meaning. Any point can be the barycenter of infinitely many sets of points, possibly using sets of weights. We refer the reader to appendix A for a formal description of the notion of a barycenter.

We further note that in order to obtain meaningful distances these values must be scale-invariant. This means that the distance between points *P* and *Q* must be the same as the distance between the points *P* and *cQ*, where *c* is a strictly positive number. Indeed: the total output of a research group can be several orders of magnitude larger than that of one expert. This difference must not play a role in determining cognitive distances. The barycenter method explained above and in particular formulae (1) satisfy this requirement as multiplying all $m_j$s with the same strictly positive factor leads to the same barycenter.

As stated earlier, we also used another quantitative approach, which was referred to as a barycenter approach in N dimensions. In this approach, we used a matrix of similarity values between the WoS SCs as made available by Rafols et al. (2010) at http://www.leydesdorff.net/overlaytoolkit/map10.paj. These authors created a matrix of citing to cited SCs based on the Science Citation Index (SCI) and Social Sciences Citation Index (SSCI), which was cosine-normalized in the citing direction. The result is a symmetric N×N similarity matrix (here, N=224) which we denote by $S = (s_{ij})_{ij}$. Now each unit's publications are represented by an N-dimensional vector. Coordinates of these vectors are the number of publications in each WoS SC. Then we wrote in (Rahman et al., 2015):

> *A barycenter in N dimensions is determined as the point $C = (C_1, C_2, \ldots, C_N)$, where:*
>
> $$C_k = \frac{\sum_{j=1}^{N} m_j s_{jk}}{T} \qquad (2)$$
>
> *Here $s_{jk}$ denotes the k-th coordinate of WoS subject category j, $m_j$ is the number of publications in subject category j, and $T = \sum_{j=1}^{N} m_j$ is the total number of publications.*

Observe that we replaced (for clarity) *L = A* as used in Rahman et al. (2015) by *S* (the similarity matrix) and *M* (in the original publication) by *T* (the total number of publications of the unit under investigation). In Rahman et al. (2015) we provided concrete calculations of distances between these so-called barycenters of units. Although formula (1) and (2) look the same, their interpretation is different as will be explained.

The numerator of formula (2) is equal to the *k*-th coordinate of $S * M$, the multiplication of the similarity matrix S and the column matrix of publications $M = (m_j)_j$. We next include an example showing what is actually happening.

Let *N* be 4. Assume that a unit has publication column $M = \begin{pmatrix} 4 \\ 1 \\ 0 \\ 0 \end{pmatrix}$.

Let $S = \begin{pmatrix} 1 & 0.1 & 0.3 & 0.8 \\ 0.1 & 1 & 0.2 & 0.1 \\ 0.3 & 0.2 & 1 & 0.6 \\ 0.8 & 0.1 & 0.6 & 1 \end{pmatrix}$ then $S * M = \begin{pmatrix} 4.1 \\ 1.4 \\ 1.4 \\ 3.3 \end{pmatrix}$.

Dividing by *T*=5 yields the vector $\frac{1}{5} \begin{pmatrix} 4.1 \\ 1.4 \\ 1.4 \\ 3.3 \end{pmatrix} = \begin{pmatrix} 0.82 \\ 0.28 \\ 0.28 \\ 0.66 \end{pmatrix}$.

Clearly, the resulting column vector is not a barycenter as it is not obtained as the result of a barycenter operation on a set of vectors.

The column vector $S * M / T$, resulting from the matrix product of matrix *S* and column vector *M/T*, can be interpreted as a 'pseudo-normalized' publication vector that takes similarity into account. It is not a real normalization because normalization has been performed with respect to the sum of the coordinates of *M* and not with respect to $S * M$. For this reason, we call $S * M$ a similarity-adapted publication vector, denoted as $M_{sa}$. In this example, this means that, for instance, the one publication in the second category also contributes (for 10%) to the publications in category 1. Although there is no original publication in category 4, we end up with a value 3.3 because category 1 and category 4 are very similar (80% similarity) and also the second category contributed. If we neglect similarity then *S* is the identity matrix and publication columns stay unchanged.

Hence, the distances we calculated through the N-dimensional approach in Rahman et al. (2015) are not normalized and not scale-invariant although they should be. In retrospect, we admit that referring to the N-dimensional approach as 'barycenters in N dimensions' was a misnomer. Therefore, we suggest that in Rahman et al. (2015), any reference to N dimensions and in particular to calculations of barycenters in N dimensions, in the paper itself as well as in the supplementary online materials should be ignored.

## Appendix A. Barycenters

A barycenter is the result of an operation performed on a set of vectors. Let $X = (X_n)_{n=1,\ldots,k}$ be a set of vectors in m-dimensional space, $\mathbf{R}^m$. Then its barycenter $B_X$ is the result of the following mapping:

$$B: (\mathbf{R}^m)^k \to R^m : (X_n)_{n=1,\ldots,k} \to B_X = \frac{1}{k}\sum_{n=1}^{k} X_n \qquad (3)$$

An example: let m = 2, k = 4 and $X_1 = \begin{pmatrix}0\\0\end{pmatrix}, X_2 = \begin{pmatrix}0\\1\end{pmatrix}, X_3 = \begin{pmatrix}2\\1\end{pmatrix}$ and $X_4 = \begin{pmatrix}2\\0\end{pmatrix}$. Then the barycenter of this set of four vectors is: $\frac{1}{4}\left(\begin{pmatrix}0\\0\end{pmatrix} + \begin{pmatrix}0\\1\end{pmatrix} + \begin{pmatrix}2\\1\end{pmatrix} + \begin{pmatrix}2\\0\end{pmatrix}\right) = \frac{1}{4}\begin{pmatrix}4\\2\end{pmatrix} = \begin{pmatrix}1\\0.5\end{pmatrix}$. This is the standard barycenter of the set of vertices $X_1$, $X_2$, $X_3$ and $X_4$ of a rectangle in the plane. More generally, one may assign a positive weight to each vector. If $m_n$ is the weight assigned to vector $X_n$ then the (generalized) barycenter (or center of gravity) is the result of the following mapping:

$$B: (\mathbf{R}^+, R^m)^k \to R^m : (m_n, X_n)_{n=1,\ldots,k} \to B_x = \frac{1}{T}\sum_{n=1}^{k} m_n X_n \qquad (4)$$

where $T = \sum_{n=1}^{k} m_n$. If all weights are set equal to 1 then one recovers formula (3).

Clearly, any vector can be the barycenter of infinitely many sets of vectors and weights. This is the main reason why the term 'barycenter' has no meaning on its own. In an extremely formal way, one may even say that any vector $X_0$ is the barycenter of itself, by taking the set of vectors equal to the singleton set {X} and weight equal to 1.